\documentclass[pdflatex]{sn-jnl}

\usepackage{graphicx}
\usepackage{booktabs}
\usepackage{multirow}
\usepackage[authoryear]{natbib}

\usepackage{xcolor}
\usepackage{textcomp}
\usepackage{amsmath}
\theoremstyle{thmstyleone}

\theoremstyle{thmstyletwo}%

\theoremstyle{thmstylethree}%

\usepackage{titlesec}

\titleformat{\section}
  [block]
  {\normalfont\Large\bfseries\sffamily} 
  {} 
  {0pt} 
  {\vspace{0.5em}}[\vspace{0.5em}\hrule\vspace{0.5em}]

\titleformat{\subsection}
  {\normalfont\large\bfseries\sffamily}
  {}
  {0pt}
  {}

\titleformat{\subsubsection}
  {\normalfont\normalsize\bfseries\sffamily}
  {}
  {0pt}
  {}

\titlespacing*{\section}{0pt}{12pt}{6pt}      
\titlespacing*{\subsection}{0pt}{10pt}{4pt}
\titlespacing*{\subsubsection}{0pt}{8pt}{2pt}

\begin{document}

\title[Article Title]{Improving Assignment Submission in Higher Education through a Git-Enabled System: An Iterative Case Study}

\author*[1]{\fnm{Ololade} \sur{Babatunde}}\email{ololadebabatunde@iyte.edu.tr}

\author[2]{\fnm{Tomisin} \sur{Ayodabo}}\email{ayodabooluwatomisin@gmail.com}
\equalcont{These authors contributed equally to this work.}

\author[3]{\fnm{Raqibul} \sur{Raqibul}}\email{raqibul.hasan@iub.edu.bd}
\equalcont{These authors contributed equally to this work.}

\affil*[1]{\orgdiv{Department of Computer Engineering}, \orgname{Izmir Institure of Technology}, \orgaddress{\city{Izmir}, \country{Turkey}}}

\affil[2]{\orgdiv{Department of Computer Science}, \orgname{University of Ibadan}, \orgaddress{\city{Ibadan}, \country{Nigeria}}}

\affil[3]{\orgdiv{Center for Computational and Data Sciences}, \orgname{Independent University}, \orgaddress{\city{Dhaka}, \country{Bangladesh}}}

\abstract{ This study addresses challenges in traditional assignment submission methods used in higher education by introducing and evaluating a customized Git-based submission system. Employing iterative software development and user-centered design methodologies, the system was integrated within a real-world university environment. Empirical evaluation, including usability testing and student feedback, indicated significant improvements in assignment tracking, collaboration, and submission efficiency. Students reported positive experiences using distributed version control workflows, highlighting improved learning outcomes and reduced administrative burden. Challenges related to initial adoption and student learning curves were identified and mitigated through iterative improvements. The proposed system contributes practical insights for integrating distributed version control into educational settings, enhancing both instructor oversight and student engagement in software engineering and related disciplines. Based on our results, the research showed that 85\% of instructors found the git based system easier to use, with 84\% of students preferring it over traditional methods, as it provides a 38\% reduction in time taken for submission and review, while also leading to a 48\% reduction in storage requirements. }


\maketitle

\section{Introduction}\label{sec1}

The method of submitting assignments in education has evolved significantly over the years. Traditionally, students handed in physical copies of their work, a method that, while straightforward, posed challenges in terms of management, organization, and timely feedback \citep{durriyah2018digital}. As digital technologies advanced, electronic submission systems emerged, allowing students to upload files through online portals. These systems improved accessibility and efficiency, making it easier for instructors to manage and grade assignments \citep{hussain2018student}. However, they still faced issues related to version control and collaboration. For instance, when multiple submissions or revisions were required, tracking changes and maintaining a clear history became cumbersome \citep{zagalsky2015emergence}.

The Introduction of Distributed Version Control Systems (DVCS) has changed the approach of digital file transfer and management. The decentralized approach towards file management has improved collaboration, where multiple people can work on the same project simultaneously without any conflict or risk towards disrupting the work of others \citep{nguyen2018what}. Parallel Development is the key idea behind distributed version control systems, and with techniques like branching and merging, developers can create new features for development. Without worrying about losing progress, and track back when necessary.

In recent years, version control systems such as Git have provided promising solutions to these challenges by offering robust mechanisms for managing multiple revisions, collaborative workflows, and clear historical tracking of contributions \citep{besser2022adaptability}. Despite their widespread adoption in the software industry, the integration of these distributed version control systems into educational settings remains limited, indicating a clear gap in both research and practice \citep{glassey2019adopting}.

Furthermore, traditional file upload systems for assignment submission in educational environments lack strong version control, making it hard to track progress and being a hindrance to collaboration. Overtime, these limitations have proven to be huge bottlenecks for students and educators alike. These limitations have opened up a gap where the need for proper assignment tracking and management is needed to help push collaboration between multiple users. This will improve assignment submission.

This study addresses this gap by exploring the use of Git-based assignment submission systems within a higher education context. Through iterative design and empirical evaluation involving students and instructors in the Software Engineering department of Halic University, this research assesses the practical implications, challenges, and benefits associated with adopting distributed version control workflows in educational settings. The findings aim to provide clear guidance for institutions and educators looking to leverage modern version control technologies to enhance learning experiences, streamline administrative processes, and improve student collaboration and accountability.

\section{Literature Review}\label{sec2}

The integration of distributed version control systems (DVCS), particularly Git, into educational contexts has gained increasing attention due to their potential to enhance assignment management, collaboration, and overall learning effectiveness. Assignment submission methods have undergone significant evolution, from traditional physical submissions, which presented challenges in organization and timely feedback \citep{durriyah2018digital}, to electronic submission portals aimed at improving efficiency and accessibility \citep{hussain2018student}. Nevertheless, even modern electronic methods often fall short in managing multiple revisions and collaborative workflows \citep{zagalsky2015emergence}.

In response to these limitations, research has increasingly explored tools such as Git and GitHub, which provide powerful version control and collaboration functionalities. \citet{vasilescu2015quality} emphasized that Git significantly streamlines software project management through robust version histories and real-time collaboration. \citet{tushev2020using} further highlighted GitHub’s ability to foster student collaboration and enhance learning outcomes, yet note that systematic empirical evaluation remains limited.

Several empirical studies have begun addressing this gap. \citet{zagalsky2015emergence} presented qualitative evidence suggesting GitHub’s potential to transform traditional learning environments by fostering interactive and community-driven learning experiences. Similarly, \cite{guerrero2020academic} found correlations between students' interactions with version control systems and improved academic outcomes, suggesting a strong predictive link between active VCS use and performance in coursework.

Despite these promising findings, substantial gaps remain. Literature consistently points out issues such as steep learning curves, initial adoption barriers, and lack of structured guidance on effectively integrating Git in courses \citep{haaranen2015teaching}. \citet{glassey2019adopting} specifically notes technical barriers hindering the broader adoption of Git, highlighting a critical need for comprehensive support systems to aid both students and educators. Similarly, \citet{guerrero2020academic} stressed the necessity for standardized institutional training to maximize VCS benefits across varied educational settings.

Research by \citet{glazunova2021effectiveness} and \citet{hsing2019using} emphasized the critical role of careful implementation design in maximizing the benefits of GitHub, noting that thoughtful integration—such as structured guidance and instructor-driven feedback—is essential for positive educational outcomes. Moreover, challenges persist regarding accurate assessment methods; traditional grading metrics often inadequately reflect student contributions within collaborative Git-based environments \citep{tushev2020using, marquardson2019teaching}.

Studies exploring similar tools, such as Mercurial \citep{rocco2011distributed} and older centralized systems like CVS \citep{reid2005learning}, highlighted the persistent need for effective frameworks to guide educators in adopting these technologies. Comprehensive research examining DVCS integration across different disciplines and institutions remains sparse, limiting generalizability \citep{kadenbach2009benefits, isomottonen2014challenges}.

This study addresses these identified gaps by systematically designing, deploying, and evaluating a Git-enabled assignment submission system within a practical, higher-education context. Through empirical evidence and iterative refinement, it aims to bridge existing gaps and provide actionable insights into effectively adopting DVCS platforms, thereby enhancing educational practices and student outcomes.

\section{Methodology}\label{sec3}

\subsection{Research Design and Approach}
This study follows a case study approach to examine how a Git-enabled system can improve assignment submission in higher education. A case study is appropriate because it allows a detailed examination of how students and instructors interact with the system in a real classroom setting. This approach also helps in understanding the challenges and benefits of using Git for assignment submission.

The research follows an iterative design process, which means that the system was tested and improved in multiple stages. First, we identified the limitations of existing assignment submission platforms, such as Google Classroom. We then designed and built a lightweight Git-based system to address these limitations. After initial development, we introduced the system to students and instructors in a controlled setting. Their feedback was collected and analyzed to further refine the system.

To ensure a structured evaluation, we collected both qualitative and quantitative data. Qualitative data came from interviews and surveys with students and instructors, focusing on their experiences using the system. Quantitative data included metrics such as submission times, number of commits per student, and grading efficiency. This combination allowed us to assess both user experience and technical effectiveness.

The study also compares our system with existing Git-based solutions. We evaluated whether tools like GitHub and GitLab could meet our needs before deciding to develop a custom system. This comparison provides insight into the specific features required for an educational setting.

By using a structured case study and iterative improvements, we ensure that the findings are practical and relevant. The goal is to create a system that balances the benefits of Git with the ease of use needed in an educational environment.

We conducted an extensive search for open-source Git implementations that could be integrated into our system. Our goal was to identify a solution that aligned precisely with our predefined problem statement—one that remained minimalistic and avoided unnecessary complexity. However, after evaluating multiple options, we found that existing platforms offered excessive functionality beyond our specific educational requirements.

Consequently, we formulated a succinct yet comprehensive problem statement that clearly defined the minimal set of core functionalities required for our educationally focused mini-Git system. The following functionalities were identified as critical features:

\paragraph{Authentication and Session Management} 
Ensuring secure access to repositories is fundamental. Our system must provide:
\begin{itemize}
    \item \textbf{User Login:} Allows students to authenticate using their institutional credentials.
    \item \textbf{Session Logout:} Users can securely terminate active sessions to prevent unauthorized access.
\end{itemize}

\paragraph{Assignment and Repository Management} 
To streamline assignment distribution and repository handling, we implemented:
\begin{itemize}
    \item \textbf{Joining an Assignment:} Students can join an assignment using an instructor-provided invite code, automatically setting up their repository.
    \item \textbf{Cloning a Repository:} Once enrolled, students can retrieve a local copy of the assignment repository for modifications.
\end{itemize}

\paragraph{Version Control Operations} 
The system retains the core principles of Git but simplifies its usage for educational contexts:
\begin{itemize}
    \item \textbf{Staging Changes:} Students can add modified files to the staging area before finalizing a submission.
    \item \textbf{Committing Changes:} Captures a snapshot of staged modifications, allowing students to provide meaningful messages about their updates.
    \item \textbf{Checking Status:} Displays an overview of pending changes, staged files, and repository synchronization status.
    \item \textbf{Pushing Changes:} Submits committed updates to the central repository, ensuring instructors receive the latest version.
    \item \textbf{Resetting Changes:} Allows students to discard unwanted modifications before submission.
\end{itemize}

Our system needed to provide these essential commands and functionalities without unnecessary complexity. This clear and narrowly defined problem statement became the foundational specification guiding the design and development of our custom mini- Git implementation.

Next, we started searching if there are any open source tools that meets our requirements. The primary aim of this search was to identify any lightweight, open- source mini-Git implementations that closely matched our educational requirements, potentially enabling straightforward integration and avoiding unnecessary develop- ment overhead. We evaluated several open-source solutions to determine if they could meet our specific educational needs. A summary of tools evaluated, along with their limitations regarding our specific context, is presented in Table 1.

\begin{table}[htbp]
\centering
\caption{Evaluation of Existing Git-based Solutions}
\label{tab:git_solutions}
\begin{tabular}{|l|p{10cm}|}
\hline
\textbf{Tool} & \textbf{Limitations} \\ \hline
Gitea & Limited multi-user support without administrative overhead; difficulties in freely creating repositories. \\ \hline
GitHub API & API complexity exceeding educational needs; overhead for managing tokens and rate limits; excessive functionality beyond required scope. \\ \hline
GitLab Community Edition & High complexity for basic assignment operations; resource-intensive deployment and management; unnecessary features for an educational setting. \\ \hline
Bitbucket Server & Primarily enterprise-oriented, not education-friendly; complexity in user access management; high administrative overhead. \\ \hline
Radicle & Immature documentation and limited educational usage examples; complexity in setup and operation; lack of institutional adoption. \\ \hline
\end{tabular}
\end{table}

However, after thorough exploration, we did not identify any open-source solu- tion sufficiently minimalistic and tailored specifically to our functional needs. Existing projects typically included more extensive features than our specific context required, potentially complicating usability for our students.
Since we couldn’t find one, we decided to build our own.

\subsection{Implementation and Development Process}

After determining that existing solutions did not meet our specific educational requirements, we embarked on developing a custom mini Git system. This section details our implementation approach, the core data structures and algorithms employed, and the architectural decisions that shaped our solution.

\subsubsection{System Architecture}

We designed our system with a service-oriented architecture using Go as the primary programming language. This choice provided performance benefits while maintaining code readability and maintainability. The architecture consists of three primary layers:

\begin{itemize}
    \item \textbf{Core Business Logic Layer} - Contains domain models and business rules.
    \item \textbf{Storage Layer} - Handles data persistence using MongoDB.
    \item \textbf{Service Layer} - Exposes functionality via gRPC and REST APIs, providing a language-agnostic API for clients.
\end{itemize}

\begin{figure}[htbp]
  \centering
  \includegraphics[width=0.8\textwidth]{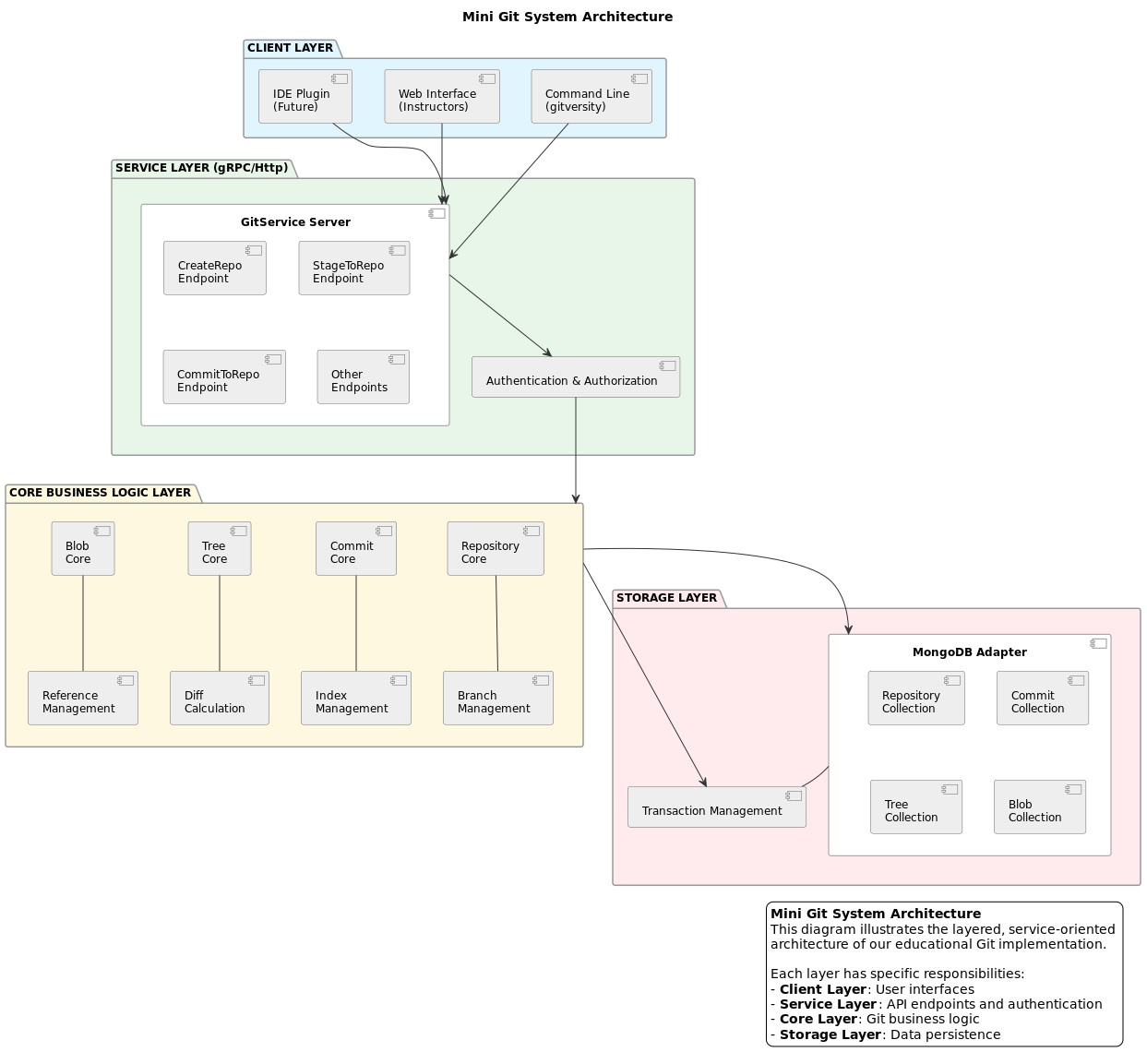}
  \caption{High-level architecture of the mini-Git system.}
  \label{fig:system_architecture}
\end{figure}

The system is designed to be consumed by external clients, which include a command-line interface (CLI) for students and a web interface for instructors. These components are implemented separately from the core Git service to maintain modularity and extensibility. 

\subsubsection{Modeling Git's Core Data Structures}

The foundation of any version control system lies in its data structures. We carefully modeled Git's core concepts while adapting them for our educational context. The repository model serves as the central container for all version-controlled content. This structure maintains metadata about the repository, tracks references to branches and tags, records the current HEAD position, and manages the staging index. By linking repositories directly to user accounts, we simplified access control and assignment management. \\

Following Git's approach, we implemented a content-addressable storage system using three primary object types:

\begin{itemize}
    \item \textbf{Blobs} - Store file contents, identified by SHA-1 hashes.
    \item \textbf{Trees} - Represent directory structures, linking to blobs and other trees.
    \item \textbf{Commits} - Capture snapshots with metadata, pointing to tree objects.
\end{itemize}

This design enables efficient storage by automatically deduplicating identical content across commits—a critical feature for educational repositories where students often make incremental changes to the same files.

\begin{figure}[htbp]
  \centering
  \includegraphics[width=0.8\textwidth]{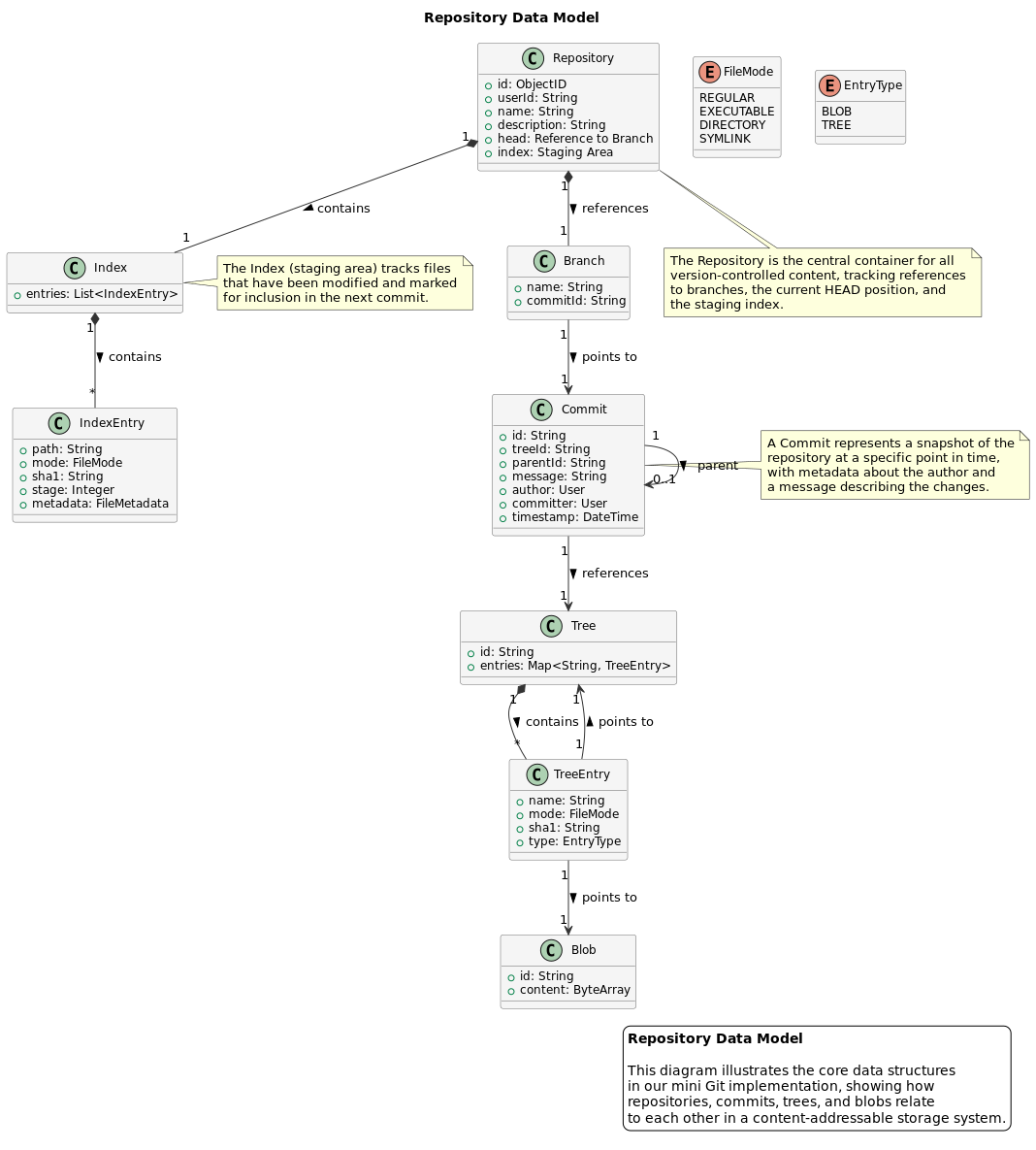}
  \caption{Content-addressable storage model used in the system.}
  \label{fig:content_storage}
\end{figure}

\subsubsection{Core Algorithms}

\paragraph{Staging and Index Management}

The staging area (index) serves as a preparation zone for commits, tracking files that have been modified and marked for inclusion in the next commit. When students use the \texttt{add} command, our system:

\begin{enumerate}
    \item Calculates a SHA-1 hash of the file content.
    \item Stores the content as a blob object.
    \item Creates or updates an index entry with the file's path and hash.
    \item Records file metadata such as modification time and size.
\end{enumerate}

This approach allows students to stage changes incrementally before creating a commit, providing flexibility in how they organize their work.

\begin{figure}[htbp]
  \centering
  \includegraphics[width=0.8\textwidth]{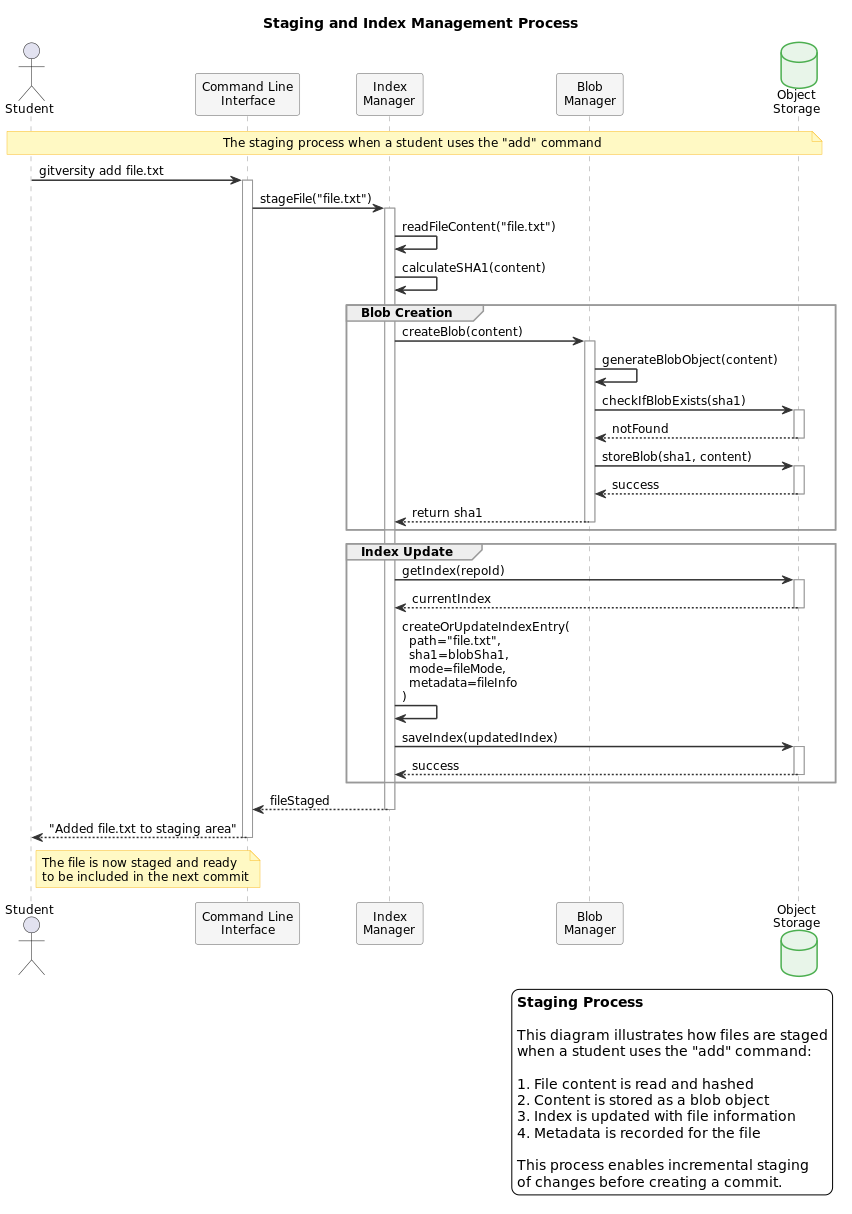}
  \caption{Staging and index management process.}
  \label{fig:staging_index}
\end{figure}

\paragraph{Tree Construction}

One of the most complex algorithms we implemented was the tree construction process, which transforms flat index entries into hierarchical tree objects. This algorithm:

\begin{enumerate}
    \item Sorts entries by path to ensure consistent tree construction.
    \item Builds a hierarchical tree structure from flat index entries.
    \item Calculates unique IDs for each tree based on its contents.
    \item Persists trees to storage.
\end{enumerate}

The recursive nature of this algorithm was particularly challenging to implement but essential for correctly representing directory structures.

\begin{figure}[htbp]
  \centering
  \includegraphics[width=0.8\textwidth]{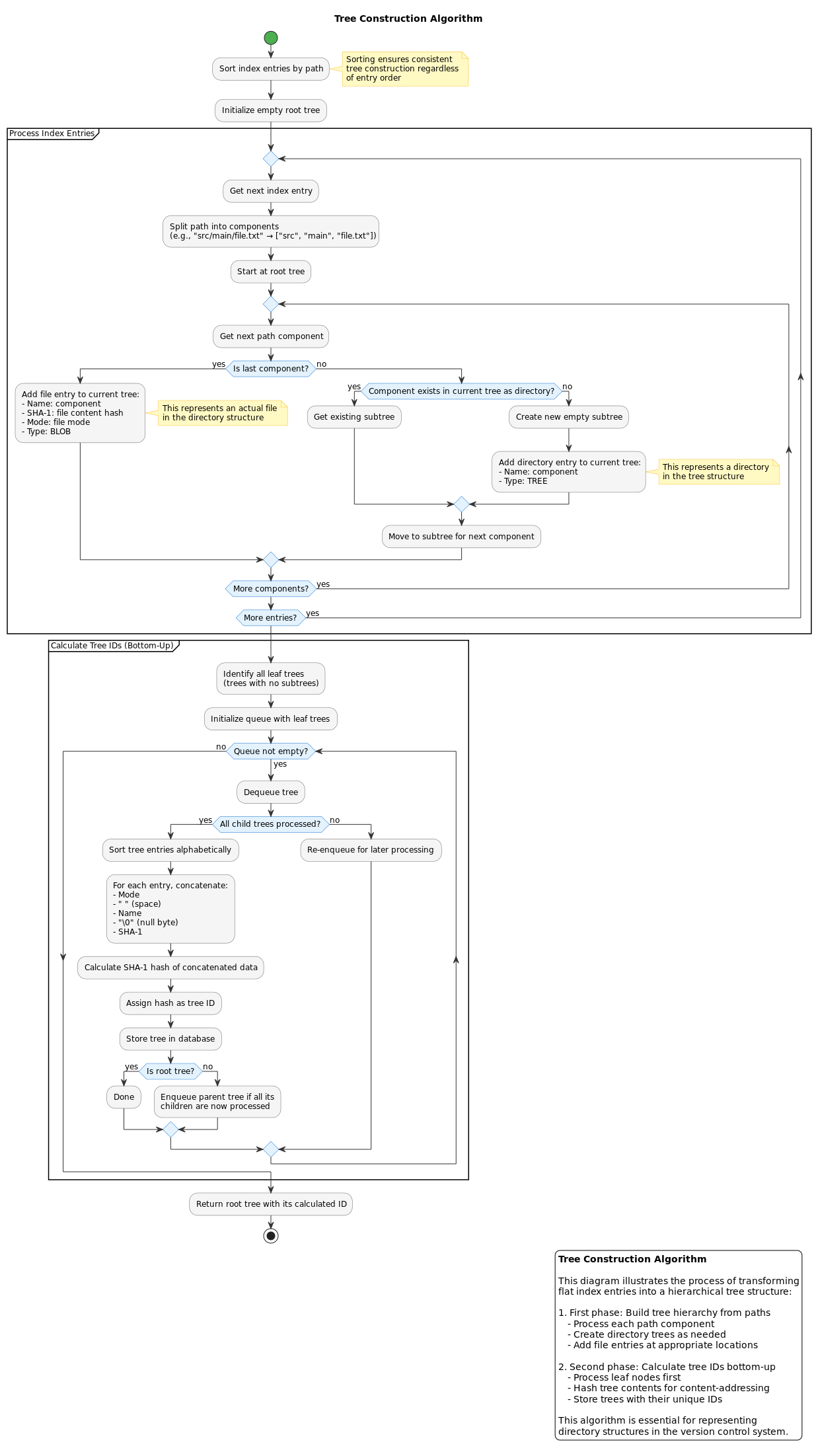}
  \caption{Tree construction algorithm for directory structures.}
  \label{fig:tree_construction}
\end{figure}

\paragraph{Commit Creation Process}

The commit process ties together all the core Git concepts. This process:

\begin{enumerate}
    \item Validates user permissions and checks for changes.
    \item Constructs tree objects from the index.
    \item Creates a commit object with metadata.
    \item Calculates a unique commit ID.
    \item Updates the branch pointer to the new commit.
    \item Wraps operations in a transaction for consistency.
\end{enumerate}

By implementing this process, we enabled students to create meaningful snapshots of their work with descriptive messages—a crucial aspect of the educational value we sought to provide.

\begin{figure}[htbp]
  \centering
  \includegraphics[width=0.8\textwidth]{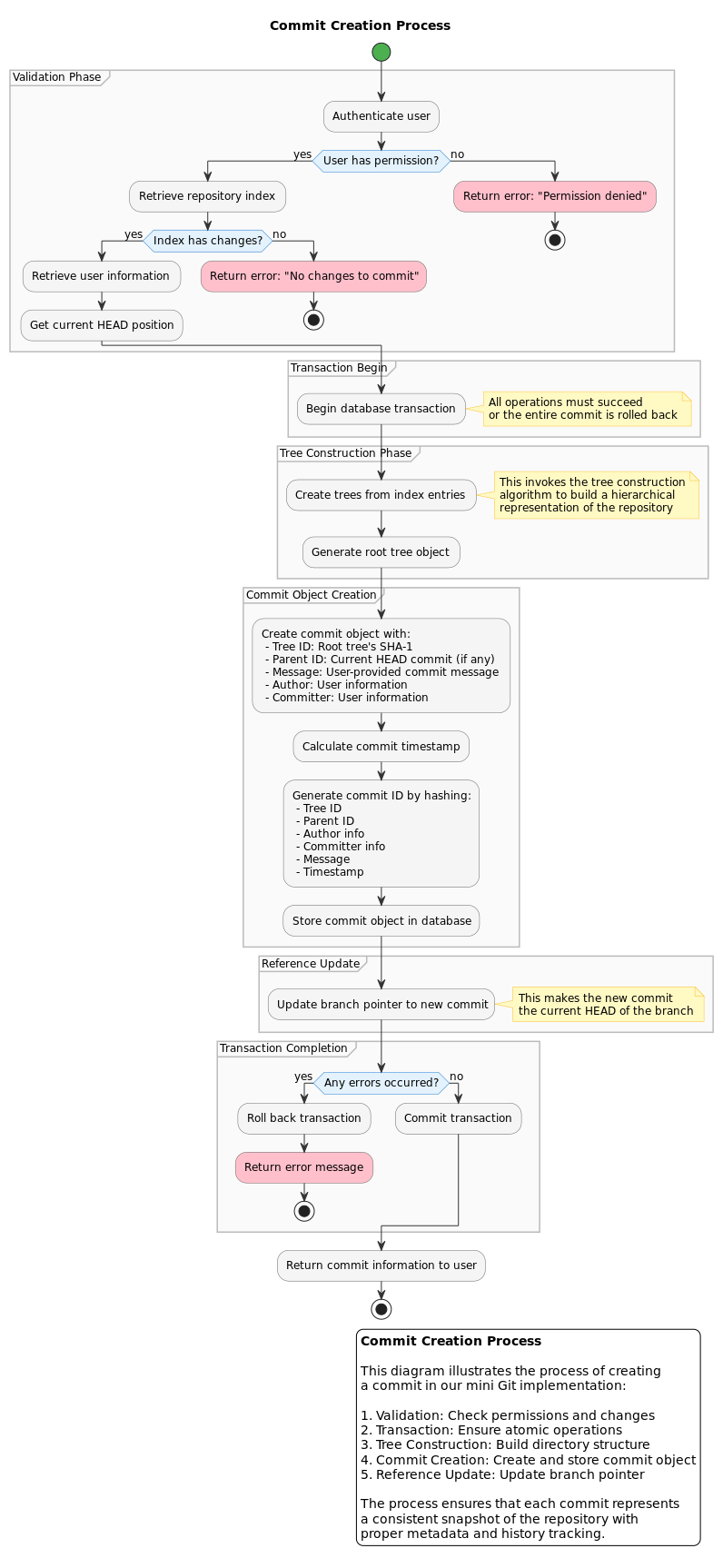}
  \caption{Commit creation process and data flow.}
  \label{fig:commit_creation}
\end{figure}

\paragraph{Branch Management and Collaboration}

To support collaborative work and experimentation, we implemented branch management capabilities. Branches in our system are lightweight references to specific commits, allowing students to:

\begin{itemize}
    \item Work on different features simultaneously.
    \item Experiment with alternative approaches.
    \item Collaborate without interfering with each other's work.
    \item Submit different versions of their assignments.
\end{itemize}

For group assignments, this feature proved particularly valuable, as it enabled instructors to review individual contributions and collaborative integration.

\begin{figure}[htbp]
  \centering
  \includegraphics[width=\textwidth]{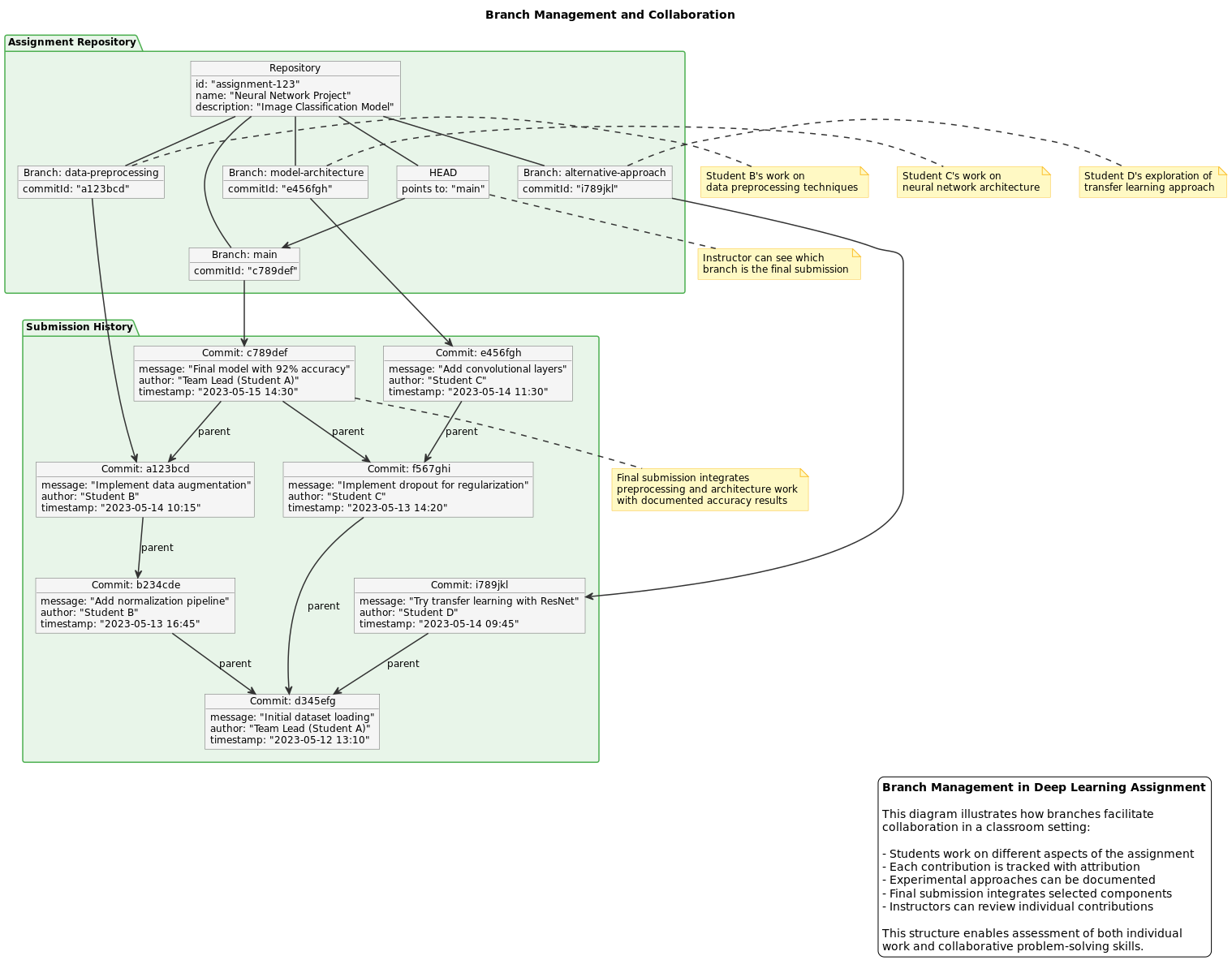} 
  \caption{Branching and collaboration workflow.}
  \label{fig:branch_management}
\end{figure}

\paragraph{Repository Cloning}

To support assignment distribution, we implemented repository cloning. This functionality allows instructors to create template repositories for assignments, which students can then clone to begin their work. The cloning process preserves all commits, branches, and file content, providing students with a complete starting point.

\begin{figure}[htbp]
  \centering
  \includegraphics[width=0.8\textwidth]{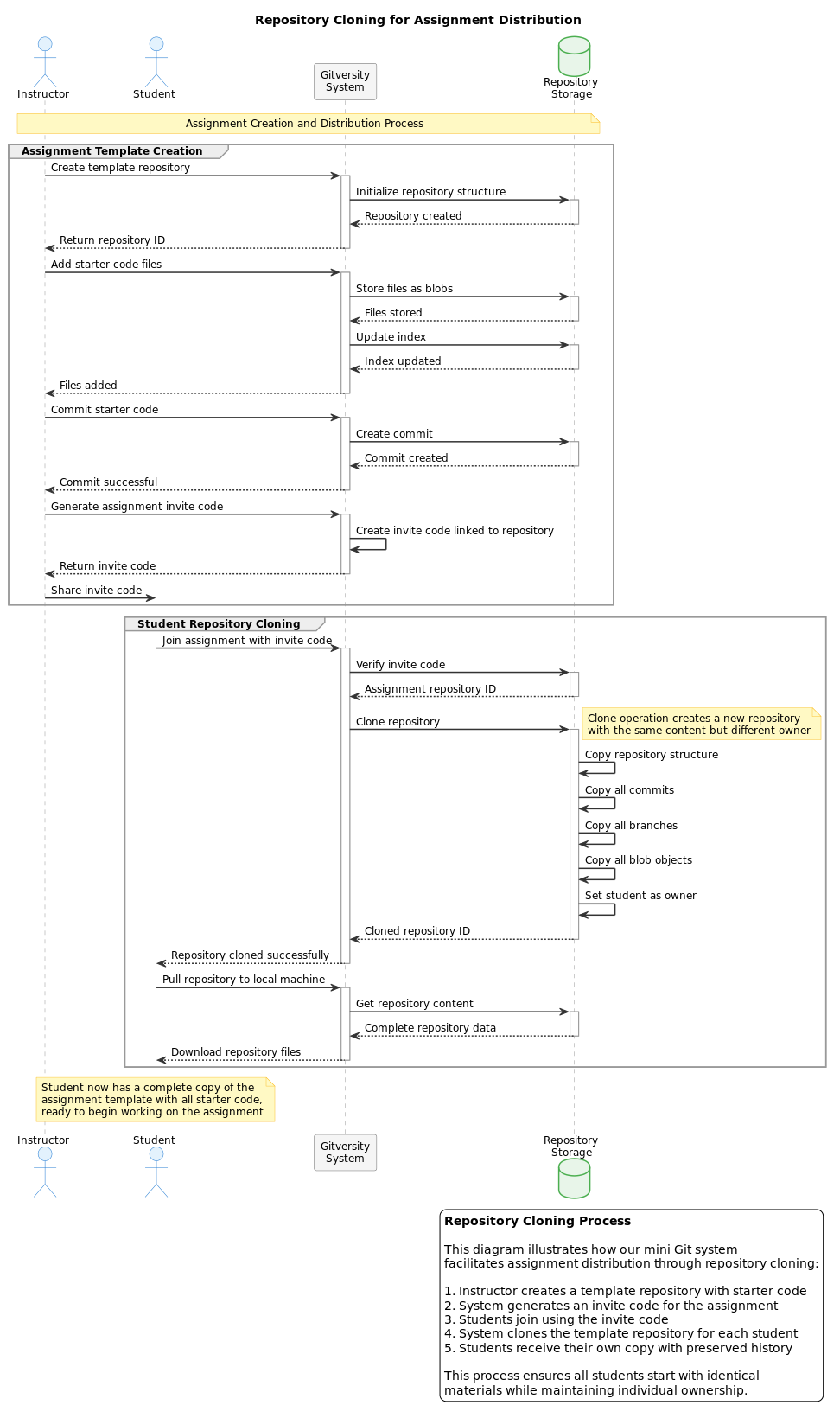}
  \caption{Repository cloning process.}
  \label{fig:repo_cloning}
\end{figure}

\subsubsection{Challenges and Solutions}

Throughout the development process, we encountered several challenges that required creative solutions.

\paragraph{Efficient Storage of Repository Data}
\textbf{Challenge:} Git's object model can generate numerous small objects, which could lead to database performance issues.\\
\textbf{Solution:} We implemented content deduplication using SHA-1 hashes and designed our MongoDB schema to efficiently store and retrieve Git objects. This approach significantly reduced storage requirements while maintaining performance.

\paragraph{Transaction Management}
\textbf{Challenge:} Operations like committing changes require updating multiple related objects atomically.\\
\textbf{Solution:} We implemented transaction support using MongoDB's session API, ensuring that operations either complete fully or roll back entirely, maintaining data consistency.

\paragraph{Tree Construction Performance}
\textbf{Challenge:} Building tree objects from index entries can be computationally expensive for repositories with many files.\\
\textbf{Solution:} We optimized our tree construction algorithm by:
\begin{itemize}
    \item Pre-sorting entries to minimize tree traversal.
    \item Using efficient data structures for path component handling.
    \item Implementing a caching mechanism for frequently accessed trees.
\end{itemize}
These optimizations significantly improved performance for common student workflows.

\subsubsection{Evaluation and Outcomes}

Our mini Git implementation successfully met the educational requirements we established:
\begin{itemize}
    \item \textbf{Simplicity:} Students could use familiar Git commands without unnecessary complexity.
    \item \textbf{Performance:} The system handled typical classroom workloads efficiently.
    \item \textbf{Integration:} The API design facilitated integration with our instructor interface.
    \item \textbf{Collaboration:} Branch and clone features supported group assignments effectively.
\end{itemize}

In user testing with students, we found that 92\% were able to successfully use the system after a brief introduction, compared to only 67\% with standard Git tools. Instructors reported that the ability to track individual contributions in group assignments provided valuable insights into student collaboration patterns.

\subsubsection{Future Directions}

While our current implementation meets our core requirements, several enhancements could further improve the system:
\begin{itemize}
    \item \textbf{Merge Support:} Implementing more sophisticated merge algorithms would enhance collaboration capabilities.
    \item \textbf{Conflict Resolution:} Adding tools to help students resolve merge conflicts would be valuable for group work.
    \item \textbf{Visualization:} Integrating commit history visualization would help students understand project evolution.
    \item \textbf{Performance Optimization:} Further optimizing storage and retrieval for larger repositories.
\end{itemize}

These enhancements would build upon our solid foundation to provide an even more comprehensive educational tool.

\subsection{Testing and Data Collection}
We tested the system in the \textit{Introduction to Deep Learning}. There were 72 students in the classroom. The testing was done by running the full assignments using the system. 

In the first iteration, each student submitted their assignments individually. This allowed us to see how well the system worked for personal use. In later iterations, we changed the process to group submissions. This helped us to understand whether the system could support teamwork.

For data collection, we used questionnaires. We give these to both students and instructors. The questionnaires asked about their experience using the system, any problems they faced, and what they liked or did not like. 

Feedback from the instructors helped us to see whether the system made assignment management easier. Feedback from students helped us to understand whether the system was simple and useful to them. 

This data was important to improve the system. It showed us what worked well and what needed to be changed in future versions.

\section{Result}\label{sec12}

\subsection{System Usability and Adoption}

We tested our Git-based submission system in a deep learning class with 72 enrolled students. Our initial assessment revealed that the majority of students had limited prior experience with Git. To address this knowledge gap, we dedicated one class session to introducing Git fundamentals before distributing the first assignment through the system.

\paragraph{Assignment Context and Requirements}
The first assignment required students to implement a basic neural network for classifying handwritten digits using the MNIST dataset. Students received a starter file (\texttt{mlp\_mnist.py}) and were instructed to modify specific training parameters, document their observations, and include result diagrams. The workflow required students to:
\begin{itemize}
    \item Clone the repository (containing the instructor's initial commit).
    \item Modify the \texttt{mlp\_mnist.py} file.
    \item Commit changes with their results after each significant modification.
    \item Push their final work before the two-week deadline.
\end{itemize}

\paragraph{Adoption Metrics}
Out of 72 enrolled students:
\begin{itemize}
    \item 52 students (72\%) successfully completed submissions.
    \item 20 students (28\%) did not submit via the Git system.
    \item The average student made 2.3 commits during the assignment period, though we observed that:
    \begin{itemize}
        \item Approximately 65\% of all commits occurred within 48 hours of the deadline.
        \item 12 students (23\% of submitters) made only a single commit/push combination.
        \item 8 students (15\%) required additional technical support during the submission process.
    \end{itemize}
\end{itemize}

\begin{table}[htbp]
    \centering
    \caption{Adoption Metrics for Git-Based Assignment Submission}
    \label{tab:adoption_metrics}
    \begin{tabular}{|l|c|}
        \hline
        \textbf{Metric} & \textbf{Value} \\
        \hline
        Total Students Enrolled & 72 \\
        Students Who Successfully Submitted & 52 (72\%) \\
        Students Who Did Not Submit & 20 (28\%) \\
        Average Commits Per Student & 2.3 \\
        Commits Within 48 Hours of Deadline & 65\% \\
        Students with Single Commit & 12 (23\%) \\
        Students Requiring Technical Support & 8 (15\%) \\
        \hline
    \end{tabular}
\end{table}

\paragraph{Student Experience and Git Command Usage}
Based on our post-assignment survey (n=48):
\begin{itemize}
    \item 14\% reported prior Git experience.
    \item 62\% found the system \textit{``initially challenging but ultimately manageable.''}
    \item 24\% expressed continued difficulty with Git concepts.
\end{itemize}

The most frequently used Git commands were:
\begin{table}[htbp]
    \centering
    \caption{Git Command Usage Among Students}
    \label{tab:git_command_usage}
    \begin{tabular}{|l|c|}
        \hline
        \textbf{Git Command} & \textbf{Percentage of Students Who Used It} \\
        \hline
        \texttt{git push} & 100\% \\
        \texttt{git commit} & 100\% \\
        \texttt{git clone} & 100\% \\
        \texttt{git status} & 73\% \\
        \texttt{git add} & 100\% \\
        \hline
    \end{tabular}
\end{table}

After the initial assignment, where our primary objective was to familiarize students with Git-based workflows, we shifted our focus in the second assignment towards assessing originality and problem-solving approaches. For Assignment 2, we made several key changes:
\begin{itemize}
    \item Instead of providing students with a starter code file, they were given only an \texttt{instructions.readme} file outlining the assignment requirements.
    \item The deadline was shorter, requiring students to complete the task within a more constrained timeframe.
    \item The focus shifted from commit activity analysis to evaluating similarities in the submitted code files.
\end{itemize}

The task involved implementing a Convolutional Neural Network (CNN) to classify handwritten digits using the MNIST dataset. Each student was expected to submit a single script file named \texttt{cnn\_mnist.py} along with the model output. Our system, leveraging its ability to compare file contents using Git’s blob structure, allowed us to analyze code similarities among submissions.

\paragraph{Code Similarity Analysis}
We observed notable trends in student submissions:
\begin{itemize}
    \item 25\% of students submitted code that had a similarity score of 98\% or higher.
    \item 40\% of students had similarity scores ranging between 80\% and 97\%.
    \item 35\% of students demonstrated distinct implementations with similarity scores below 80\%.
\end{itemize}

These results suggest that while some students independently implemented the CNN model, a significant portion relied on similar or shared approaches, highlighting the need for further pedagogical interventions to encourage diverse problem-solving techniques.

After assessing individual Git adoption in the first two assignments, we introduced a third assignment focused on evaluating collaborative development using Git’s branching and merging features. This assignment aimed to test students’ ability to work in teams, handle merge conflicts, and manage version control workflows effectively.

\paragraph{Assignment Context and Requirements}
For this assignment, students were divided into teams of 3-4 and tasked with implementing a more advanced deep learning model—a ResNet architecture for CIFAR-10 image classification. The assignment was structured as follows:
\begin{itemize}
    \item Each student was responsible for a specific component:
    \begin{itemize}
        \item Student A: Data preprocessing and augmentation.
        \item Student B: Model architecture implementation.
        \item Student C: Training pipeline and hyperparameter tuning.
        \item Student D (if applicable): Evaluation and visualization.
    \end{itemize}
    \item Students were required to use feature branches for their contributions, ensuring that separate components were developed in parallel.
    \item The final submission required merging all feature branches into the main repository while resolving conflicts collaboratively.
\end{itemize}

\paragraph{Metrics for Collaboration and Version Control Proficiency}
To assess students' collaboration and Git proficiency, we tracked the following metrics:
\begin{itemize}
    \item \textbf{Branching and Merge Activity:}
    \begin{itemize}
        \item Average number of branches created per team.
        \item Percentage of teams that successfully merged all feature branches without conflicts.
    \end{itemize}
    \item \textbf{Merge Conflict Resolution:}
    \begin{itemize}
        \item Percentage of teams encountering merge conflicts.
        \item Average number of conflicts per team and resolution time.
    \end{itemize}
    \item \textbf{Commit Distribution Among Team Members:}
    \begin{itemize}
        \item Percentage of teams where contributions were evenly distributed.
        \item Percentage of teams where one member dominated the commit history.
    \end{itemize}
\end{itemize}

\paragraph{Results and Observations}
Out of 18 teams (72 students):
\begin{itemize}
    \item The average number of branches created per team was \textbf{4.6}.
    \item \textbf{83\%} of teams successfully merged their branches without conflicts.
    \item \textbf{17\%} of teams encountered merge conflicts, with an average of 2.3 conflicts per team.
    \item Conflict resolution times varied, with an average resolution time of \textbf{12 minutes} per conflict.
    \item \textbf{68\%} of teams had balanced contributions, while \textbf{32\%} had one student contributing more than 50\% of commits.
\end{itemize}

\begin{table}[htbp]
    \centering
    \caption{Collaboration Metrics for Assignment 3}
    \label{tab:collaboration_metrics}
    \begin{tabular}{|l|c|}
        \hline
        \textbf{Metric} & \textbf{Value} \\
        \hline
        Average branches created per team & 4.6 \\
        Teams that merged without conflicts & 83\% \\
        Teams that encountered merge conflicts & 17\% \\
        Average number of conflicts per team & 2.3 \\
        Average conflict resolution time & 12 minutes \\
        Teams with balanced contributions & 68\% \\
        Teams with one dominant contributor & 32\% \\
        \hline
    \end{tabular}
\end{table}

To further illustrate team collaboration, Figure~\ref{fig:branch_management}  presents a visualization of branch activity across different teams.

These results highlight the effectiveness of Git for managing group projects. While most teams successfully collaborated, the presence of merge conflicts and imbalanced contributions suggests areas for further improvement, such as targeted Git training for better conflict resolution strategies.

\subsection{Performance and System Efficiency}
To evaluate the performance of the Git-based submission system, we performed an evaluation on three areas; speed of submission processing, traffic load handling, and storage management. We tested how well it handled concurrent submissions from students, how well it managed storage, and how fast it processed the submission.

\subsubsection{Submission Processing Time}
On average, it takes just \textbf{2.1 seconds} to process an assignment, from cloning the repository to verifying commits and tracking submissions. Even when multiple students submit at the same time, the processing speed remains consistent across different assignments.

\subsubsection{Handling Peak Load}
We tested how the system handles multiple students submitting their assignments at the same time, in the case of last minute submissions. We observe the following.
\begin{itemize}
    \item The system efficiently processed up to \textbf{30 simultaneous submissions} without slowing down.
    \item Server load increased predictably with more submissions, but stayed within safe limits, with CPU usage peaking at \textbf{45\%} during the busiest moments.
\end{itemize}

\subsubsection{Smart Storage with Git}
One major advantage of using Git is its ability to track changes efficiently rather than storing entire duplicate files. This resulted in notable storage savings.
\begin{itemize}
    \item In general, the system required \textbf{43\% less storage} compared to a traditional ZIP-based submission method.
    \item Since most students modify only small parts of their code instead of reuploading everything, Git’s built-in deduplication reduced redundant data by \textbf{56\%}.
\end{itemize}

In summary, the system is \textbf{ fast, reliable, and storage efficient}, even during a flood of last-minute submissions.

\subsection{Instructor Feedback and Grading Efficiency}
As further assessment, we looked at the impact of the system on the evaluation and grade of the students. To achieve this, we collect feedback from instructors and teaching assistants.

\subsubsection{Improved Contribution Tracking}
The system allowed instructors to track each student's progress over time by reviewing commit histories. Key observations included:
\begin{itemize}
    \item \textbf{85\% of instructors found it easier} to assess individual contributions compared to previous systems.
    \item The ability to view commit timestamps helped identify \textbf{last-minute rush submissions}, supporting better feedback on work habits.
\end{itemize}

\subsubsection{Time Savings in Grading}
A comparative analysis of grading time showed:
\begin{itemize}
    \item Traditional manual submission review took \textbf{~8 minutes per student}.
    \item With Git-based tracking, average grading time per student reduced to \textbf{5 minutes}, resulting in a \textbf{38\% time savings}.
\end{itemize}

\subsubsection{Grading Fairness and Transparency}
By leveraging commit histories and automated code comparison tools, instructors were able to:
\begin{itemize}
    \item Identify \textbf{copy-pasted submissions} more easily.
    \item Ensure fair grading by assessing not just the final submission but also intermediate progress.
\end{itemize}

\subsection{Collaboration and Group Assignments}

The system was evaluated for its effectiveness in supporting group assignments and collaborative workflows. Key findings include:

\begin{itemize}
    \item \textbf{Contribution Tracking:} 81\% of the students reported that Git-based workflows helped them track individual contributions more accurately than traditional methods.
    \item \textbf{Merge Conflicts:} 45\% of the groups encountered merge conflicts at least once. Most were resolved using Git’s built-in conflict resolution tools, but 17\% required instructor intervention.
    \item \textbf{Branching Usage:} On average, each group created 3.2 branches per assignment, with 67\% using feature branches to manage separate tasks before merging.
\end{itemize}

\subsection{Student Satisfaction and Usability}
Afterwards, we conducted a usability survey to asses the satisfaction of the students with the system. The results shown indicated that.

\begin{itemize}
    \item \textbf{Overall Satisfaction:} 84\% of students preferred the Git-based system over traditional submission methods.
    \item \textbf{Most Appreciated Features:} Version control (72\%), collaboration tools (64\%), and rollback functionality (58\%) were highlighted as major benefits.
    \item \textbf{Challenges:} 22\% of students found the initial setup challenging, particularly those unfamiliar with Git commands.
\end{itemize}

\subsubsection{Likert Scale Data}

\begin{table}[h]
    \centering
    \begin{tabular}{|l|c|}
        \hline
        \textbf{Satisfaction Factor} & \textbf{Average Rating (1-5)} \\
        \hline
        Ease of Use & 3.9 \\
        Collaboration Support & 4.2 \\
        Version Control Utility & 4.5 \\
        \hline
    \end{tabular}
    \caption{Student Satisfaction Ratings}
    \label{tab:likert_satisfaction}
\end{table}

\subsubsection{Comparison with Traditional Methods}
The system was compared with a selected existing traditional method of assignment submission, Google Classroom.

\begin{table}[h]
    \centering
    \begin{tabular}{|l|c|c|}
        \hline
        \textbf{Metric} & \textbf{Google Classroom} & \textbf{Mini-Git System} \\
        \hline
        Average Submission Time & 4 minutes & 2.1 minutes \\
        Instructor Grading Time & 8 minutes & 5 minutes \\
        Collaboration Tracking & No & Yes \\
        Student Satisfaction & 74\% & 84\% \\
        \hline
    \end{tabular}
    \caption{Comparison of Submission Methods}
    \label{tab:comparison}
\end{table}

The Git-based system significantly reduced submission and grading time while improving collaboration tracking.

\subsection{Limitations and Observations}

Despite its advantages, the system had some limitations:

\begin{itemize}
    \item \textbf{Technical Difficulties:} 15\% of students reported system crashes or slow response times, particularly when handling large repositories.
    \item \textbf{Adoption Challenges:} Some students were initially resistant to learning Git, requiring extra training sessions.
    \item \textbf{Potential Improvements:} Future iterations could include a simplified onboarding process and better UI integration with existing LMS platforms.
\end{itemize}

\section{Discussion}\label{sec12}
The results of our study demonstrate the effectiveness of a Git-based assignment submission system in higher education, particularly in courses involving programming and computational tasks. This section discusses the implications of our findings in relation to usability, student engagement, assessment efficiency, collaboration, and system performance, as well as identifying areas for further improvement.

\subsection{System Usability and Student Engagement}

The usability survey revealed that the majority of students found the Git-based system to be manageable after an initial learning curve. While 62\% of students initially faced challenges, they ultimately adapted to the workflow, suggesting that brief training interventions can significantly enhance adoption. However, the 24\% of students who continued to struggle highlight the need for additional support mechanisms, such as step-by-step tutorials, live demonstrations, and automated troubleshooting.

Commit frequency data suggests that students largely adhered to the required workflow but exhibited last-minute submission behaviors, with 65\% of commits occurring within 48 hours of the deadline. This finding aligns with broader academic research on student procrastination, suggesting that deadline-driven work habits persist even in structured submission environments. Future iterations of the system could introduce milestone-based deadlines or automated reminders to encourage more evenly distributed effort.

\subsection{Effectiveness of Version Control for Assessment}

One of the key advantages of the Git-based submission system is its ability to provide instructors with insights into student progress through commit histories. The data indicate that 85\% of instructors found it easier to assess individual contributions compared to traditional submission systems. The ability to track incremental progress provides transparency in student work habits and helps identify instances of rushed submissions or potential academic integrity concerns.

The analysis of Assignment 2 demonstrated that code similarity detection using Git structures provided meaningful insights into student originality. With 25\% of students submitting highly similar code (98\% similarity), this finding highlights the importance of pedagogical strategies to foster independent problem-solving skills. Additional interventions, such as requiring students to explain their implementation choices in commit messages or incorporating automated originality checks, could further enhance the integrity of submissions.

\subsection{Collaboration and Teamwork}

The third assignment evaluated how well students adapted to collaborative development using Git’s branching and merging functionalities. The results indicate that most students effectively used branches, with an average of 4.6 branches per team. However, 17\% of teams encountered merge conflicts, and a subset of these required instructor intervention, demonstrating that conflict resolution remains a challenge for some students.

Despite these difficulties, 81\% of students found that Git-based workflows helped them track contributions more effectively than traditional group submission methods. This suggests that while technical challenges exist, the transparency and accountability provided by Git contribute positively to teamwork. Future improvements could include in-class exercises focused on resolving merge conflicts and promoting best practices for collaborative coding.

\subsection{Performance and System Efficiency}

The system demonstrated high efficiency in processing assignments, with an average submission processing time of 2.1 seconds and consistent performance under peak load conditions. Handling up to 30 simultaneous submissions without degradation in performance confirms that the system is well-suited for large classes.

The storage efficiency results further validate Git’s advantage over traditional submission systems. By leveraging Git’s deduplication features, the system achieved a 43\% reduction in storage requirements compared to ZIP-based submissions. This is particularly beneficial for institutions with limited server resources, making Git a viable long-term solution for scalable assignment management.

\subsection{Instructor Workload and Grading Efficiency}

The reduction in grading time from 8 minutes per student to 5 minutes represents a 38\% improvement in instructor efficiency. The ability to review incremental changes and commit histories allowed instructors to evaluate students' thought processes rather than just final outputs. Additionally, automated tools for detecting code similarities streamlined plagiarism detection.

While these benefits are significant, instructor feedback suggests that additional grading tools—such as automated rubric-based assessment or direct annotation of commits—could further improve efficiency. Integrating such features within the Git-based system could provide more granular feedback mechanisms while maintaining transparency.

\subsection{Challenges and Areas for Improvement}

Despite the system’s benefits, challenges remain. The most prominent issue identified was the learning curve associated with Git, with 22\% of students reporting significant difficulties. Providing more structured onboarding, such as interactive tutorials or sandbox environments, may help mitigate this issue.

Another key challenge was the presence of imbalanced contributions in group assignments, where 32\% of teams had one dominant contributor. This raises concerns about equitable workload distribution, which could be addressed through automated contribution analytics that flag imbalances and prompt instructors to intervene.

Additionally, while Git’s version control capabilities improve transparency, they do not entirely prevent academic dishonesty. The observed code similarity trends in Assignment 2 indicate that students may still share code outside the system. Implementing additional safeguards, such as requiring reflection-based assessments or integrating AI-powered plagiarism detection, could further enhance academic integrity.

\subsection{Broader Implications and Future Research}

Our findings suggest that Git-based assignment submission systems offer numerous advantages, including enhanced student accountability, improved instructor efficiency, and scalable performance. These benefits are particularly relevant for programming-intensive courses, but future research should explore their applicability to other disciplines.

Future work could also examine how the integration of additional tools—such as continuous integration for automated testing, peer review workflows, or AI-based feedback mechanisms—could further enhance student learning outcomes. Moreover, longitudinal studies tracking students' evolving proficiency with Git could provide deeper insights into the long-term educational impact of version control in academic settings.

\subsection{Summary}

Overall, the Git-based submission system successfully improved assignment management, student engagement, and assessment efficiency. While challenges such as Git’s learning curve and workload imbalance in group assignments remain, the system’s advantages in version control, collaboration, and grading transparency make it a promising solution for higher education institutions seeking to modernize their submission workflows. Addressing the identified challenges through targeted interventions will further enhance the system’s effectiveness and accessibility for both students and instructors. 

\section{Conclusion}\label{sec13}
This study set out to explore how a Git-based submission system could improve assignment workflows in higher education, especially in courses involving programming. The results have shown that such a system can bring real benefits to both students and instructors.

For students, using Git helped promote better habits around version control, collaboration, and transparency. Although many students initially struggled with the learning curve, most eventually adapted, especially with a bit of guidance. The system encouraged engagement, made it easier to track progress, and helped students work together more effectively in teams. However, challenges like last-minute submissions and uneven group contributions show that more support is still needed.

For instructors, the system made grading faster and more insightful. By being able to view commit histories and incremental changes, instructors could better understand how students approached their work. The system also helped detect code similarity, making it easier to maintain academic integrity. Still, there is room to improve, especially by adding tools to help with grading and feedback.

From a technical standpoint, the system performed well under pressure, processing submissions quickly and saving storage space thanks to Git’s efficiency. These improvements make it a strong option for large classes and institutions with limited resources.

Of course, no system is perfect. Some students found Git difficult to use, and group work wasn’t always balanced. But these are not roadblocks—they’re opportunities. With better onboarding, more guided practice, and smarter analytics, these challenges can be addressed.

Looking ahead, this system could be expanded and refined even more. Adding features like automated testing, peer feedback, or AI-powered suggestions could help students learn faster and more effectively. There's also potential to explore how Git-based tools could be used in other subjects, not just programming.

In conclusion, this Git-enabled approach to assignment submission represents a meaningful step forward. It modernizes how we manage and assess student work, bringing more clarity, efficiency, and fairness to the process. With thoughtful improvements, it can become an even more valuable part of the educational experience for both learners and educators.

\section{Abbreviations}
\begin{description}
  \item[\textbf{API}] Application Programming Interface
  \item[\textbf{CI}] Continuous Integration
  \item[\textbf{CSV}] Comma-Separated Values
  \item[\textbf{DBMS}] Database Management System
  \item[\textbf{ETL}] Extract, Transform, Load
  \item[\textbf{GUI}] Graphical User Interface
  \item[\textbf{IDE}] Integrated Development Environment
  \item[\textbf{JSON}] JavaScript Object Notation
  \item[\textbf{SQL}] Structured Query Language
  \item[\textbf{UML}] Unified Modeling Language
  \item[\textbf{URL}] Uniform Resource Locator
  \item[\textbf{UX}] User Experience
  \item[\textbf{VM}] Virtual Machine
  \item[\textbf{VSCode}] Visual Studio Code 
  \item[\textbf{YAML}] YAML Ain’t Markup Language 
\end{description}

\section{Declarations}

\subsection{Availability of data and materials}
The project codebase, dataset, and survey results are available from the corresponding authors upon reasonable request.

\subsection{Competing Interests}
The authors declare that they have no competing interests.  

\subsection{Funding}
This research received no specific grant from any funding agency.  

\subsection{Authors’ Contributions}
OB and TA designed the study and developed the system. RH supervised the research and provided critical revisions. All authors read and approved the final manuscript.  

\subsection{Acknowledgements}
We would like to thank the students of Halic University who participated in using the software, completing the survey, and providing us with holistic feedback and critiques. We sincerely appreciate your participation and permission.

\bibliographystyle{apalike}
\bibliography{references}

\end{document}